\documentclass[b5paper]{article}
\usepackage[12pt]{extsizes}
\usepackage[left=1.4 cm, right=1.4  cm, top= 1.5 cm]{geometry}
\usepackage{amsmath}
\usepackage{mathptmx,mathtools}
\usepackage{dirtytalk}
\usepackage{enumitem}
\usepackage{multicol}
\usepackage[colorlinks=true,linkcolor=blue,urlcolor=red]{hyperref}
\usepackage{xcolor}
\begin{document}		
\begin{center}
\textbf{Null Lagrangians in Schwarzian mechanics}\\
\vspace{0.5 cm}

Pratik Majhi and Madan Mohan Panja\\
Department of Mathematics, Visva-Bharati University, Santiniketan 731235, India\\

Pranab Sarkar\\
Department of Chemistry, Visva-Bharati University, Santiniketan 731235, India\\
and\\
Benoy Talukdar\\
Department of Physics, Visva-Bharati University, Santiniketan 731235, India\\
\end{center}
\title
\maketitle
\section*{Abstract}

In addition to standard and non-standard Lagrangians of classical mechanics, we consider, in this work,  null Lagrangians that (i) identically satisfy the Euler-Lagrange equation and at the same time can be expressed as (ii) the total derivative of some scalar function. As an addendum to the properties in (i) and (ii) we find that null Lagrangians are also characterized by (iii) vanishing energy functions or Jacobi integrals. By working with higher-order ${SL(2, R)}$  invariant Schwarzian derivatives introduced recently by Krivonos (Phys. Rev. D \textbf{109}, 065029(2024)) we demonstrate that these Schwarzians, especially the even –order ones, provide a natural basis to introduce higher-order null  Lagrangians in Schwarzian mechanics.

\begin{multicols}{2}

Physical theories are often formulated using standard Lagrangian functions [1] represented by \( L = T - V \), where \( T \) is the kinetic energy of the system under study and \( V \), the corresponding potential function. There also exist Lagrangian functions which involve neither \( T \) nor \( V \). Despite this, the sought functions lead to the equations of motion via the Euler-Lagrange equation. Consequently, these Lagrangian functions are qualified as ‘non-standard’. The non-standard Lagrangians play a role in the study of non-linear systems [2,3]. An additional family of Lagrangian functions often called the null Lagrangians (NLs) is also of considerable current interest. By a null Lagrangian, we mean one, the Euler-Lagrange equation of which vanishes identically. The relevance of NL to ordinary differential equations has been discussed by Olver [4] who clearly pointed out that the Euler-Lagrange equations \(E(NL) = 0 \) vanish identically for all values of the arguments (say \( t, x \)) of \( NL \) only if it is a total derivative of some function \( P \) of \( t, x \) and of derivatives of \( x \) with respect to \( t \). The function \( P \) is often called the gauge function. In the following, we deal with two well-known examples of null Lagrangians quoted by Musielak [5] and point out that, in addition to the well-known properties of NL quoted above, the energy functions or Jacobi integrals of NLs also vanish identically.

The first null Lagrangian of our interest is given by

 \begin{equation}
 \mathrm{NL_{1}(x , \dot{x}) = c_{1}x \dot{x}}
\end{equation}
\\
where \(c_{1}\) is an arbitrary constant and, as before, \( x \) is function of time \(t\). Here over dot denotes differentiation with respect to time \(t\). The first-order Lagrangian (1) leads to vanishing Euler-Lagrange equation with a gauge function \( P = ({\frac{1}{2 c_{1}}})x^2\). The second example of our interest is again a first-order Lagrangian

\begin{equation}
	\mathrm{NL_{2}(x, \dot{x}) = \frac{a_{1}\dot{x}}{a_{2}x + a_{4}}}
\end{equation}
\\	
Here, \(a\)'s are arbitrary constant. We have now \( E(NL_{2}) = 0\) with \( P = \frac {a_{1}}{a_{2}}|a_{2}x + a_{4}| \).\\
The Jacobi integral or the energy function of any first-order Lagrangian (\(L^{(1)}\)) is given by

\begin{equation}
\mathrm{J(L^{(1)}) = x \frac{\delta{L^{(1)}}}{\delta{\dot{x}}} - L^{(1)}} 
\end{equation}
\\
Using Eqs. (1) and (2) successively in Eq. (3) we find that \( J (NL_{1}) = J (NL_{2}) = 0\). This indicates that the essential properties of a null Lagrangian, namely, \(E(NL) = 0\) and that \(NL\) is the gradient of a scalar function should be supplemented by simultaneous vanishing of its Jacobi integral.\\
The properties and applications of \( NLs\) have been explored in different areas of physics and mathematics [6,7]. The construction procedure of null Lagrangians have also been studied in the literature [8] with particular emphasis on their use in the calculus of variations. The cases of both first and higher-order Lagrangians were considered. The object of the present work is to demonstrate that the null Lagrangian has a natural space in non-relativistic conformally  invariant systems.

For an arbitrary function \(q(t)\)the expression  

  \begin{equation}
 	\mathrm{\lbrace q(t), t \rbrace = \frac{\dddot{q}(t)}{\dot{q}(t)} - \frac{3}{2} \left( \frac{\ddot{q}(t)}{\dot{q}(t)} \right)^2}
 	\end{equation} 
\\ 	
defines the well known Schwarzian derivative [9]. It is invariant under fractional transformation and allows one to construct the \( SL(2, {R}) \)-invariant action functional

  \begin{equation}
  \mathrm{S = 1/2\int dt\ \lbrace q(t), t \rbrace}
  \end{equation}
\\
such that any function of \( \lbrace q(t), t \rbrace \) can be used as Lagrangians to determine the equation of motion of a higher-derivative \( SL(2, {R}) \)-invariant system. One dimensional mechanics so defined is often called the Schwarzian mechanics[10].

The Schwarzian derivative (4) can be written in an alternative form

\begin{equation}
\mathrm{ \lbrace q(t), t \rbrace = \frac{d}{dt} \left(\frac{\ddot{q}(t)}{\dot{q}(t)}\right) - \frac{1}{2} \left(\frac{\ddot{q}(t)}{\dot{q}(t)}\right)^2},
\end{equation}
\\
where \( \frac{\ddot{q}(t)}{\dot{q}(t)} \) is called the pre-Schwarzian derivative. While the Schwarzian derivative is invariant under \( SL(2, {R}) \) transformation \(q(t) = \frac{af(t) + b}{cf (t) + d}\), with \( a, b, c, d \) real or complex constants such that \( ad - bc = 0 \), the pre-Schwarzian is not. Interestingly, if one chooses to work with a Lagrangian defined by the square of the pre-Schwarzian derivative, namely,

\begin{equation}
{L^{(2)} = {\frac{1}{2}}\left(\frac{\ddot{q}(t)}{\dot{q}(t)}\right)^2},
\end{equation}
\\
one arrives at the following results[11].\\ 
The second-order energy function \(J\left( L^{(2)} \right) = \ddot q(t) \frac{ dL^{(2)}}{d\ddot{q}} - L^{(2)} + \dot q(t)  \left(\frac{ dL^{(2)}}{d\dot q(t)} - \frac{d}{dt}\left(\frac{ dL^{(2)}}{d{q(t)}}\right) \right)\) becomes equal to  -\(\lbrace q(t), t \rbrace \) and while the equation of motion is obtained as a fourth-order nonlinear differential equation

\begin{equation}
\mathrm{ q^{(4)}(t) \dot{q}(t)^{2} - 4 \dot{q}(t)\ddot{q}(t) \dddot{q}(t) +  3\ddot{q}^{(3)}(t) = 0}
\end{equation}
\\
The result in (8) has been obtained from the higher-order Euler-Lagrange equation[12]

\begin{equation}
ELeq = \sum_{i=0}^n (-1)^i \frac{d^i}{dt^i} \frac{\partial L}{\partial q^{(i)}}
\end{equation}
\\
for \( n = 2 \). In this context, it appears rather interesting to envisage a similar study for a third-order Lagrangian defined by the Schwarzian derivative. The energy function of our third-order Lagrangian can be obtained from the general expression

\begin{equation}
J(L) = \sum_{r=1}^n q^{(r)}(t)\sum_{k=0}^n r (-1)^k \frac{d^k}{dt^k} \frac{\partial L}{\partial q^{(r+k)}(t)}
\end{equation}
\\
for the Jacobi integral of the \( n^{th} \)-order Lagrangian. Interestingly, we find \( J(L^{(3)}) =  \lbrace q(t), t \rbrace \) and an equation of motion which is exactly the same as that in (8). Thus it is natural to compute results for equations of motion and Jacobi integrals using Lagrangians represented by higher-order Schwarzian derivatives and thus  gain some physical insight from them. We shall see in the course of our study this will provide an opportunity to look for null Lagrangians in Schwarzian mechanics.

Higher-order generalization of the Schwarzian derivative in (4) has been considered by a number of authors. For example, more than fifty years ago Aharonov [13] provided definitions of higher-order analogues of \( \lbrace q(t), t \rbrace \). Relatively recently, Tamanoi  [14] introduced another set of higher-order Schwarzians. In close analogy with Aharonov invariants, Schippers [15] defined higher Schwarzians \( S_{n} \lbrace q(t) \rbrace \) inductively by means of the recurrence relation

\begin{equation}
\mathrm{ S_{n+1} (q(t))  = \frac{d}{dt}S_{n} (q(t))  - (n-1) \frac{\ddot{q}(t)}{\dot{q}(t)}S_{n} (q(t))} 
\end{equation}
\\
\( n\geq 3\), with \( S_{3} (q(t)) = \lbrace q(t), t \rbrace \). The higher-order Schwarzians obtained from (11) is not invariant under \( SL(2, {R}) \)  transformation. The same is true for other such existing results.

In a recent publication Krivonos [16] proposed a physical model for the origin of higher Schwarzians and treated them as Goldstone fields associated with the generators of the Virasoro algebra [17]. This provided  a new set of \( SL(2, {R}) \) invariant higher Schwarzians. We denote them by \( \sigma , s \) rather than \( z, s \) as in ref.16 such that \( \sigma_{3} = \lbrace q(t), t \rbrace\) the Schwarzian derivative while \( \sigma_{4}, \sigma_{5} \) etc stand for higher Schwarzians. In the following we   choose these higher Schwarzians from ref.16 as Lagrangian functions and compute the Euler-Lagrange equation and Jacobi integrals by using  (9) and (10) respectively. For \( L^{(4)} = \sigma_{4} \) we found 

\begin{equation}
{ELeq = 0} \hspace{0.3 cm} and \hspace{0.3 cm} {J(L^{(4)}) = 0}
\end{equation}
\\
with the Gauge function  \( P = \lbrace q(t), t \rbrace \). Thus  the higher Schwarzian \( \sigma_{4}\) represents  a null Lagrangian.
 
Let us now consider the next higher-order Lagrangian given by 
\( L^{(5)} = \sigma_{5}\). In this case we get a sixth-order nonlinear equation of motion

\begin{equation}
\begin{split}
\mathrm{{\dot q(t)^{4} q^{(6)}(t) - 6 \dot{q}(t)^{3}\ddot{q}(t){q^{(5)}}(t)}} 
- 10\dot{q}(t)^{3}\dddot{q}(t)\\{q^{(4)}}(t) + {\frac{45}{2}}\dot{q}(t)^{2}\ddot q(t)^{2}{q^{(4)}}(t) + 30\dot{q}(t)^{2}\ddot{q}(t)\\ {{\dddot q(t)^{(2)}} - 60 \dot{q}(t)^{3}\ddot{q}(t){q^{(5)}}(t)} 
+ {\frac{45}{2}}\ddot{q}(t)^{5} = 0
\end{split}
\end{equation}
\\
with the Jacobi integral
\begin{equation}
\begin{split}
\mathrm{ J(L^{(5)})  = \frac{45\ddot q(t)^4} {4\dot q(t)^4} - \frac{25\ddot q(t)^2 \dddot q(t)}{4\dot q(t)^3}} \\ {+ \frac{5\dddot q(t)^2}{\dot q(t)^2} + \frac{10\ddot q(t)\dddot q^{(4)}(t)}{\dot q(t)^2} - \frac{2{q^{(5)}(t)}}{\dot{q(t)}}},
\end{split}
\end{equation}
\\
with \( q^{(n)}(t) = \frac{d^{n} q(t)}{dt^{n}} \). Clearly’ \( L^{(5)}\)is not a null Lagrangian although \(\sigma_{5}\) is \( SL(2, {R}) \) invariant.  The reason behind  is that \( L^{(5)}\) does not follow from a Gauge function. But working with a sixth-order Lagrangian defined by \( L^{(6)} = \sigma_{6}\) we found

\begin{equation}
{ELeq = 0} \hspace{0.3 cm} and \hspace{0.3 cm} {J(L^{(6)}) = 0}
\end{equation}
\\
and with the gauge function

\begin{equation}
\begin{split}
\mathrm{ P = \frac{{q^{(5)}(t)}}{\dot{q}(t)} - \frac{5\ddot{q}(t){q^{(4)}(t)}}{\dot{q}(t)^2} - \frac{5\dddot{q}(t)^2}{\dot{q}(t)^2}} \\ {+ \frac{20\ddot{q}(t)^2\dddot{q}(t)}{\dot{q}(t)^3} - \frac{45\ddot{q}(t)^4}{4\dot{q}(t)^4}}
\end{split}
\end{equation}
\\
In this way we could demonstrate that all even-order higher Schwarzians of Krivonos[16] represent null Lagrangians while the odd-order ones are non-null.

We conclude by noting that   fist-order  null Lagrangians corresponding to both standard  and non-standard Lagrangians  have been extensively treated  in the literature with a view to look for their usefulness in physics [18,19]. This  is , however, not the case with higher-order null Lagrangians, Thus it remains an interesting curiosity to look for the role of higher-order null  Lagrangians and/or  of  higher Schwarzians of  Krivonos  in physical theories.

\section*{Acknowledgement}
One of the authors (PM) would like to thank University Grants Commission for a research fellowship.

\section*{References}
\begin{enumerate}[label={[\arabic*]}]

\item H. Goldstein, C. P. Poole and  J. L. Safko, Classical Mechanics (3rd Edition) (Addision-Wesley, San Francisco, C. A. 2002). 

\item J. F. Carinena, M. F. Ranada and M. Santander, Lagrangian formalism for nonlinear second-order Riccati equation : one-dimensional integrability and two-dimensional superintegrability,  J. Math. Phys. 46 , 062703 (2005).

\item Aparna Saha and Benoy Talukdar, Inverse vatiational problem for nonstandard Lagrangians, Rep. Math. Phys.73,  299 (2014).

\item P. J. Olver, Applications of Lie Group to Differential Equations, Springer-Verlag, New York(1993).

\item Z. E. Musielak, Nonstandard Null Lagrangians and Gauge functions for Newtonian Law of Inertia, Physics 3, 903 (2021).

\item J. –M. Levy-Leblond. Comm. Math. Phys.12, 64 (1969).

\item G. Saccomandi and R. Vitolo, J. Math. Sciences 136, 4470 (2006).

\item P. J. Olver and J. Sivaloganathan, The structure of null Lagrangian,  Nonlinearity 1,389 (1988).

\item V. Ovsienko and S. Tabachnikov,  What is the Schwarzian derivative? Not. Am. Math. Soc . 56, 34 (2009).

\item A. Galajinsky, Schwarzian Mechanics via nonlinear realization, Phys. Lett. B 795,277 (2019).

\item W. Krynski, The Schwarzian derivative and Euler-Lagrange equations, J. Geometry and Physics 182, 104656 (2022).

\item R. Courant and D. Hilbert, Methods in Mathematical Physics, Vol. ! (Wiley Eastern Pvt. Ltd. , New Delhi, 1975).

\item D. Aharonov, A necessary and sufficient condition condition for univalence of a meromorphic  function, Duke Math. J. 36 , 599 (1969). 

\item H. Tamanoi, Higher-order Schwarzian operators  and combinatorics of the Schwarzian derivatives, Math. Ann. 305 , 127 (1996).

\item E. Schippers, Distortion theorems for higher order Schwarzian derivatives  of univalentfunctions,Proc.Am.Math.Soc.18,3241(2000).

\item S.Krivonos.    Phys.Rev.D109,065029(2024)

\item P. Goddard and D. Olive, Kac-Moody and Virasoro algebra in relation to quantum physics, Int. J. Mod. Phys.A1,303(1986).

\item A.L. Segovia, L. C. Vestal and Z.E. Musielak, Nonstandard Null Lagrangians and Gauge functions and Dissipative forces in Dynamics, arXiv:2202.02447v2 [math-ph]7Oct.2022.

\item R. Das and Z.E. Musielak, New Role of Null Lagrangians in Derivation of equations of Motion for Dynamical Systems, arXiv:2210.09105v2 [math –ph] 30 Oct 2022.

\end{enumerate}
\end{multicols}
\end{document}